\documentclass[preprint2]{aastex62}

\usepackage{color}

\graphicspath{{./}{figure_new/}}

\received{}
\revised{}
\accepted{}
\shorttitle{Sample article}
\shortauthors{Kubota et al.}
\begin{document}

\title{Discovery of Red-shifted He-like Iron Absorption Line from Luminous Accreting Neutron Star SMC X-1}

\author[0000-0001-7298-3243]{Megu Kubota}
\affil{Department of Physics, Tokyo University of Science, 1-3 Kagurazaka, Shinjyuku-ku, Tokyo 162-0825, Japan}
\affil{High Energy Astrophysics Laboratory, RIKEN Nishina Center, 2-1- Hirosawa, Wako, Saitama 351-0198, Japan}
%\email{megu@crab.riken.jp}

\author{Hirokazu Odaka}
\affiliation{Department of Physics, The University of Tokyo, 7-3-1 Hongo, Bunkyo-ku, Tokyo 113-0033, Japan}
%\collaboration{(AAS Journals Data Scientists collaboration)}

\author{Toru Tamagawa}
\affiliation{Department of Physics, Tokyo University of Science, 1-3 Kagurazaka, Shinjyuku-ku, Tokyo 162-0825, Japan}
\affiliation{High Energy Astrophysics Laboratory, RIKEN Nishina Center, 2-1- Hirosawa, Wako, Saitama 351-0198, Japan}
%\nocollaboration

\author{Toshio Nakano}
\affiliation{High Energy Astrophysics Laboratory, RIKEN Nishina Center, 2-1- Hirosawa, Wako, Saitama 351-0198, Japan}
%\nocollaboration



\begin{abstract}

%SMC X-1 is a high-mass X-ray binary that has a luminous accreting neutron star (NS) 
%with a luminosity of $5 \times 10^{38}~\mathrm{erg\;s^{-1}}$, which is mildly super-Eddington.
%This object should give us an insight to link between classical high-luminosity NSs and recently 
%discovered ultra-luminous X-ray sources that show pulsations.
We have analyzed X-ray data of SMC X-1 obtained with \textit{Suzaku}, 
and discovered a combination of an absorption line and an underlying broadened emission line 
centered at 6.4 keV in an observation performed on May 19, 2012.
This absorption line is centered at $6.61_{-0.03}^{+0.02}$~keV with an absorption strength of $5.9_{-1.4}^{+1.7}$, 
naturally interpreted as an He$_{\alpha}$ resonance line of Fe at 6.7~keV 
that has a redshift of $4000_{-1300}^{+1400}\;\mathrm{km\;s^{-1}}$.
Although \textit{Suzaku} observed this system ten times during 11 months in 2011--2012, 
the absorption feature has been seen only in a single observation when the NS was in a rising phase of the super-orbital modulation, 
which can be regarded as an egress from occultation by an extended accretion disk.
We therefore attribute the line to a low density, highly ionized absorber in an accretion disk corona arising from the disk illuminated by the NS's intense X-rays.
This interpretation also agrees with a discussion on the photoionization degree and the line depth.

\end{abstract}



% ===========================================
\section{Introduction} \label{sec:intro}
% ===========================================
Accreting neutron stars (NSs) in high-mass X-ray binaries (HMXBs) provide us with unique laboratories 
for studying accretion physics in the presence of a strong magnetic field of $\sim 10^{12}$--$10^{13}\;\mathrm{G}$.
The accretion flow onto such a magnetized NS is channeled into the magnetic poles of the NS, 
creating shocked hot spots which are observed as X-ray pulsation \citep{Lamb_1973}.
Dynamics of the accretion flow is still poorly understood 
because of the extremely complicated physical conditions 
dominated by the strong magnetic field and by the intense X-ray radiation.
It is therefore of great importance to understand the dynamics 
and the radiation of the accretion flow as a function of the X-ray luminosity, 
which is proportional to the mass accretion rate.
Conventionally, those X-ray pulsars have been classified into low-luminosity systems 
with typical X-ray luminosities of $10^{36}\;\mathrm{erg\;s^{-1}}$ 
and high-luminosity systems with $10^{38}\;\mathrm{erg\;s^{-1}}$ \citep{vanJaarsveld_2018}.
Vela X-1, an archetypal low-luminosity accreting pulsar, 
shows strong time variability \citep{Odaka_2013, Odaka_2014}, 
probably as a result of capturing of stellar wind 
from the donor star and of formation of an unstable, temporal accretion disk.
On the other hand, high luminosity systems have persistent disks, 
which are observationally characterized by more stable luminosity as seen in Cen X-3 \citep{Muller_2011}.

An important recent discovery relevant to the accretion onto NSs is the detection of X-ray pulsation 
from a few ultra-luminous X-ray sources (ULXs), 
revealing that a part of ULXs is explained by NSs that accrete matter at super-Eddington rates \citep{Bachetti_2014}.
The X-ray luminosities of these ULX pulsars range from $10^{40}$ to $10^{41}\;\mathrm{erg\;s^{-1}}$, 
making themselves distinct from the classical high-luminosity X-ray pulsars.
Thus, it should be questioned whether there is any essential physical difference 
between the super-Eddington pulsars and the normal high-luminosity pulsars below the Eddington limit.
We could imagine many factors that make the difference: 
circumstellar environment, dynamics of the accretion disk at relatively large radii, 
and/or the columnar accretion flow approaching the magnetic pole.

We select SMC X-1, which belongs to the Small Magellanic Cloud (SMC), to approach the question.
SMC X-1 is the most luminous HMXB pulsar 
in the category of the classical high-luminosity pulsars \citep{Price_1971}, 
emitting an X-ray luminosity of $\sim 5\times10^{38}\;\mathrm{erg\;s^{-1}}$.
This binary system consists of a NS with a mass of $1.06M_\odot$ \citep{vanderMeer_2007} 
and a B0 Ib supergiant Sk 160 \citep{Schreier_1972} with a mass of $17.2M_\odot$ \citep{Reynolds_1993}. 
This object exhibits X-ray pulsation with a period of 0.71 s \citep{Lucke_1976}, 
orbital modulation with a period of 3.89 days \citep{Schreier_1972}, 
and super-orbital modulation \citep{Gruber_1984}
with a period of 40--65 days (Trowbridge et al. \citeyear{Trowbridge_2007}; Hu et al. \citeyear{Hu_2011}).
This super-orbital modulation has been interpreted as extinction 
by a precessing accretion disk \citep{Wojdowski_1998, Hu_2013} though it is still not well understood.
SMC X-1 is sufficiently bright for detailed spectral analysis 
at an X-ray flux of $1\times 10^{-9}\;\mathrm{erg\;s^{-1}\;cm^{-2}}$, 
its luminosity is mildly super-Eddington, and having eclipse 
gives us good estimates of the binary properties.
Thus, this system should provide us with a link to the ULX pulsars 
and great insight into physics of luminous accreting NSs.

SMC X-1 has been observed with \textit{Suzaku} \citep{Suzaku} ten times 
at various phases both of the orbital and of the super-orbital modulations during 11 months in 2011--2012.
We discovered a combination of a broad emission line and an absorption line at iron K-shell band around 6.5 keV in one of the ten observations. 
In this article, we focus solely on this interesting line feature only seen in that single observation.
Detailed analysis of the other observations with \textit{Suzaku} will be separately reported in future.
The present paper reports a spectral analysis of the iron line structure at 6.5 keV, 
and discusses the astrophysical origin of the absorption line 
which is related to the structure and the dynamics of the accretion flow.

% ===========================================
\section{Observation and Data Processing} \label{sec:data_reduc}
\begin{figure}[htb]
\plotone{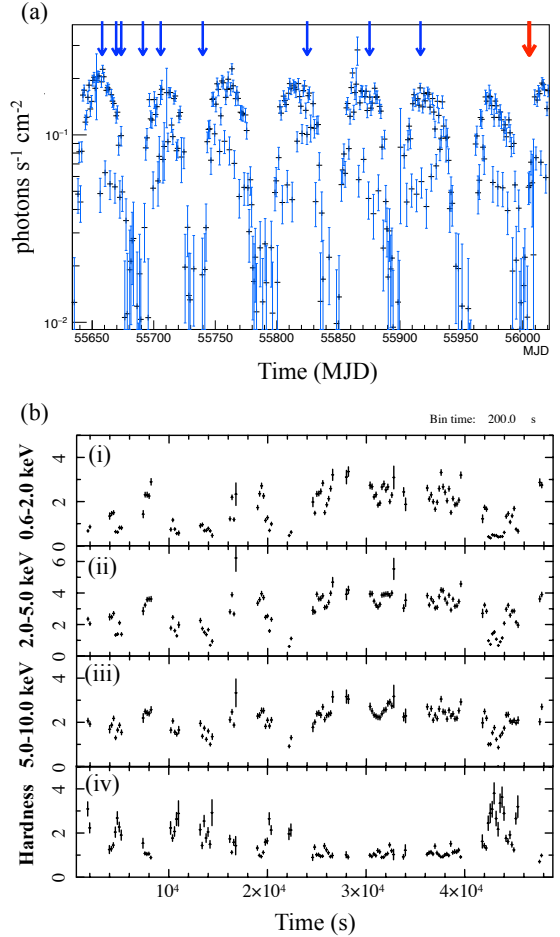}
\caption{
(a) Light curve of SMC X-1 obtained with \textit{MAXI} 
in an energy range of 2.0--20.0 keV between MJD 55635 and 55615. 
The arrows indicate the ten observations 
with \textit{Suzaku}, and this work analyzes the last red one.
(b) Background-subtracted light curves in the energy bands of 
(i) 0.6--2.0~keV, (ii) 2.0--5.0~keV, and (iii) 5.0--10.0~keV 
and (iv) hardness ratio between  5.0--10.0~keV and 0.6--2.0~keV.
\label{fig:lc}}
\end{figure}

% ===========================================
In 2011--2012, \textit{Suzaku} performed ten monitoring observations 
(PI: Neilsen) of SMC X-1 for 11 months at various phases of the orbital 
and super-orbital modulations, and each exposure time was about 20~ks.
To indicate the epochs of the \textit{Suzaku} observations, 
we show a light curve of SMC X-1 obtained with \textit{MAXI} (Matsuoka et al.\ 2009) in Figure~\ref{fig:lc} (a).
In this paper, we focus on the tenth (Obs.\ ID: 706030100; exposure time: 18.6~ks) of the total ten observations
as this only shows a significant iron absorption line.
The super-orbital modulation is seen in the light curve;
the observation was conducted on May 19, 2012 (MJD 56005) 
during rising from the dim phase of the super-orbital modulation.
The observation also corresponded to an orbital phase of 0.5 (opposite of the eclipse), 
where we used binary ephemeris of the orbital period Porb = 3.89229090(43) day
and the epoch $T_{\rm 0}$ = 55656.60 MJD (mid-eclipse), 
which was determined by the \textit{MAXI} light curve.

We reprocessed the data using \textit{HEASoft} (v6.18; Arnaud \citeyear{Arnaud_1996})
and CalDB released on June 7, 2016.
We used the data of X-ray Imaging Spectrometer (XIS, Koyama et al. \citeyear{Koyama_2007}) 
onboard \textit{Suzaku}; we analyzed data of XIS 0 and 1 
as these two were operated in the normal mode 
but XIS 3 was operated in the timing mode which might have degraded spectral performance.
To extract spectra, we accumulated the on-source events within a circle of $2'$ radius and the background events 
from a rectangle region next to the source region.
Effects of photon pileup on the spectral analysis were negligible; 
we confirmed that results of the spectral analysis did not change 
even though removing the central region where the pileup could be more significant. 
We applied resolution binning to the extracted spectra so that each bin width becomes $\sim 60$~eV. 

Figure~\ref{fig:lc} (b) shows background-subtracted light curves in three energy bands.
These light curves show a short-term variation, 
especially in low-energy bands. 
As shown in Figure~\ref{fig:lc}b (panel iv), 
the hardness increased as the count rate decreased. 
This suggests that the large time variation in the light curves are due to absorption.
Since the light curve in the highest band (5--10 keV), 
where we are interested in the iron K lines, 
does not have large variability, we analyze a time-averaged spectrum over the entire observation for this study to have sufficient photon statistics. 
In deed, we did not find any significant spectral variability 
during the observation except the degree of the absorption at low energies. 

% $$$$$$$$$$$$$$$$$$$ figure 1 $$$$$$$$$$$$$$$$$$$$
% $$$$$$$$$$$$$$$$$$$ figure 1 $$$$$$$$$$$$$$$$$$$$

% ===========================================
\section{Analysis and Results} 
\label{Sec. 3}
% ===========================================

We evaluated the combination of the emission and absorption lines at 6.5 keV.
We used \textit{XSPEC} (v12.9.1; Arnaud \citeyear{Arnaud_1996}) for this analysis.
An energy range of 3--11 keV was used since an energy band 
below 3~keV is largely contaminated by emission lines from the companion's stellar wind illuminated by the NS radiation.

\begin{figure}[htb]
\plotone{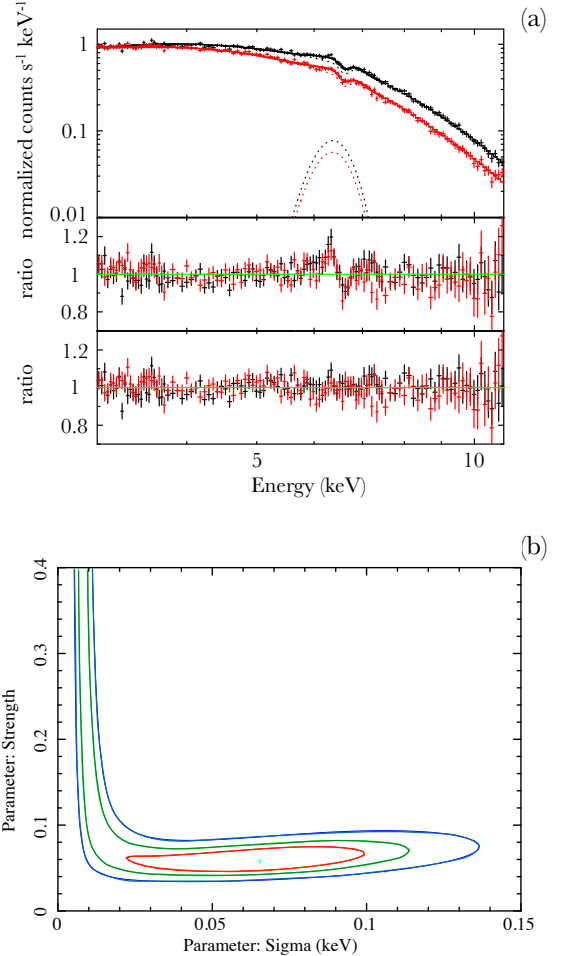}
\caption{
(a) X-ray spectra of SMC X-1 obtained with \textit{Suzaku}. 
Top panel shows the best fit spectrum. 
Middle panel show the ratio when fitted by the model of {\tt phabs*powerlaw}.
Bottom panel is ratio when the Gaussian and absorption model are employed. 
(b) Confidence level contours that show correlation 
between the line strength and width. 
Red, green, and blue show 1, 2, 3$\sigma$, respectively. 
\label{fig:spec}}
\end{figure}

First, we considered a continuum model that describes radiation of the accretion column 
to which absorption by interstellar and circumstellar media are applied.
In the soft X-ray band, below 11 keV in this study, the X-ray radiation 
from the accretion column is well described by a single power law with a typical photon index of $\sim$1.0.
Preliminarily, we fitted the continuum with an \textit{XSPEC} model of \texttt{constant*phabs*powerlaw}.
Results of this fitting are shown in Figure~\ref{fig:spec} (a).
As seen in the plot of the data-to-model ratio (the middle panel), 
we found line features composed of a broad emission and a narrow absorption 
around 6.5~keV in both the statistically independent spectra of XIS 0 and 1.
The best-fit power-law index and the hydrogen column density were $1.21 \pm 0.03$ 
and $(3.3 \pm 0.3) \times 10^{22}$~cm$^{-2}$, respectively.
The fit was not acceptable with the reduced chi-squared $\chi_\nu{}^2 = 1.45$ 
for the degree of freedom $\nu=234$, which corresponds to a null hypothesis probability of $8.2 \times 10^{-6}$.

Then, we evaluated the line feature found in the spectrum.
For this purpose, the continuum model was modified by a broad emission line and a narrow absorption line: 
\texttt{constant*phabs* \{gabs*powerlaw + Gaussian\}}.
The absorption with a Gaussian profile (\texttt{gabs}) is parameterized by the energy of the line center, its width, and the line normalization (line strength).
Similarly, the emission line is parameterized by the line center, its width, and the line flux.
Among these parameters, we fixed the line center of the broad emission line to 6.4 keV 
as the neutral iron line. 
Alternatively, we tried a 6.7 keV line as He-like iron but it did not fit to the line feature.

The best-fit spectra are shown in Figure~\ref{fig:spec} (a) and the obtained parameters are listed in Table \ref{tab:params}.
We obtained a reasonable fit with $\chi_\nu{}^2 = 1.05$ for $\nu=230$ and the null hypothesis probability is 0.28, 
which can be evaluated to be statistically acceptable, 
and the residuals between the data and the model that considered only the continuum around 6.5 keV were diminished.
The equivalent width of the Gaussian emission was determined to be $0.13_{-0.03}^{+0.04}$~keV within the confidence interval of 90\%.
The Gaussian width was $0.37_{-0.12}^{+0.13}$~keV, which is even larger than the energy resolution of XIS (130~eV at 6~keV), 
clearly showing the presence of the line broadening in the accretion flow.
The absorption energy and line strength were obtained to be $6.61 \pm 0.03$~keV and $5.9_{-1.4}^{+1.7}$, respectively.

% $$$$$$$$$$$$$$$$$$$ figure 2 $$$$$$$$$$$$$$$$$$$$
% $$$$$$$$$$$$$$$$$$$ figure 2 $$$$$$$$$$$$$$$$$$$$

% $$$$$$$$$$$$$$$$$$$ table 1 $$$$$$$$$$$$$$$$$$$$
\setcounter{table}{0}
\begin{table}[h!]
\renewcommand{\thetable}{\arabic{table}}
\centering
\caption{Best-fit model parameters} \label{tab:params}
\begin{tabular}{lccc}
\tablewidth{0pt}
\hline
\hline
Component & Parameter & value \\
\hline
% \decimals
{\tt phabs}     & $N_{\rm H}~(10^{22}~{\rm cm^{-2}})$   & $3.0 \pm 0.3$ \\
\hline
{\tt gabs}      & Energy (keV)                          & $6.61 \pm 0.03$ \\
                & width (keV)                           & $(6.6_{-5.1}^{+3.6})\times 10^{-2}$ \\
                & strength                              & $(5.9_{-1.4}^{+1.7})\times 10^{-2}$ \\
\hline
{\tt powerlaw}  & $\Gamma$                              & $1.20 \pm 0.03$ \\
                & norm                                  & $(3.1 \pm 0.2)\times 10^{-2}$ \\
\hline
{\tt Gaussian}  & Energy (keV)                          & 6.4 (Fixed) \\                      
                & width (keV)                           & $0.37_{-0.12}^{+0.13}$ \\
                & norm                                  & $(4 \pm 1)\times 10^{-4}$ \\
\hline
{\tt Constant}  & factor                                & $1.00 \pm 0.01$\\
Luminosity\tablenotemark{a}   & $L_{\rm X}~$(erg s$^{-1}$)    & $1.1 \times 10^{38}$   \\
                                                 & $\chi_\nu{}^2 $                      &  1.05(230)     \\
\hline
\end{tabular}
\tablenotetext{a}{Assuming the distance of 63~kpc without propagating its uncertainty.}
\end{table}
% $$$$$$$$$$$$$$$$$$$ table 1 $$$$$$$$$$$$$$$$$$$$

To check the significance of the absorption line, 
we estimated statistical errors of the line strength of the line for various confidence levels 
that correspond to containment probabilities of 1, 2, 3, and 5 standard deviation of the Gaussian distribution.
The obtained confidence intervals are listed in Table 2.
Even at the confidence level at $5\sigma$, the lower limit is sufficiently large against zero.
Thus, we conclude that the absorption line is clearly detected.

Figure 2 (b) shows correlations between the line strength and the line width. There are apparently two branches. The horizontal branch containing the best-fit values ranges from a narrow line width to a somewhat broadened width ($\sim 0.1\;\mathrm{keV}$) with an almost constant line strength, suggesting that the spectral resolution of the XIS and the current data statistics did not allow us to determine the line width with sufficient precision, but the line strength was relatively well determined. However, the presence of the vertical branch implies that the line strength can be much larger when the line is assumed to be narrow. This means that we can not reject the possibility of saturation of the line, though the horizontal branch (no line saturation) is marginally favored.

% $$$$$$$$$$$$$$$$$$$ table 2 $$$$$$$$$$$$$$$$$$$$
\setcounter{table}{1}
\begin{table}[h!]
\renewcommand{\thetable}{\arabic{table}}
\centering
\caption{Confidence intervals of the absorption line} \label{tab:decimal}
\begin{tabular}{lcc}
\tablewidth{0pt}
\hline
\hline
Component & \multicolumn{2}{c}{Strength ($10^{-2}$)}   \\
 & min & max \\
\hline
% \decimals
$1 \sigma$          &   5.06            &  6.93 \\
$2 \sigma$          &   4.69            &  7.38 \\
$3 \sigma$          &   4.42            &  7.74 \\
$5 \sigma$          &   4.02            &  8.33 \\
\hline
\end{tabular}
\end{table}
% $$$$$$$$$$$$$$$$$$$ table 2 $$$$$$$$$$$$$$$$$$$$

% ===========================================
\section{Discussion} 
\label{Sec. 4}
% ===========================================

We here summarize our analysis results of the \textit{Suzaku} spectra of SMC X-1 taken on March 19, 2012, when the NS was located at the orbital phase of 0.5 (opposite of the eclipse) and at the egress of the super-orbital dimming probably due to extinction by an extended precessing accretion disk \citep{Wojdowski_1998}.
The spectra showed an absorption line at $6.61 \pm 0.03$ keV, 
which can be naturally attributed to a resonance line at 6.70 keV of Fe He$_{\alpha}$ that has a redshift of $4000_{-1300}^{+1400}~\mathrm{km\;s^{-1}}$, 
and an underlying broadened emission line centered at 6.4 keV.
The line strength of this absorption line is $5.9_{-1.4}^{+1.7}$, 
suggesting that there is a substantial amount of helium-like ions of iron ionized by the NS's intense radiation on the line of sight.
The emission line seems to come from neutral atoms or mildly ionized ions of iron since the line center is consistent with a neutral iron line at 6.4 keV.
In this section we discuss the astrophysical origin of the absorption line though the broadened emission line will be investigated in a separate paper together with all other observations.

\begin{figure}[htb]
\plotone{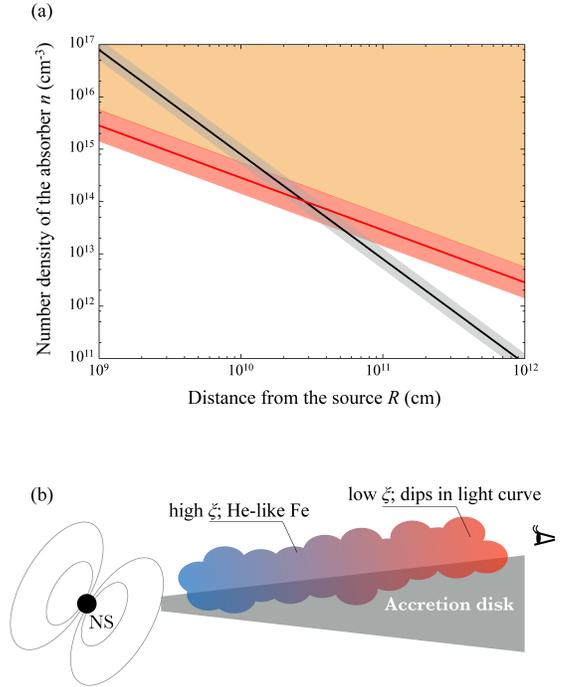}
\caption{(a) constraints of the number density and the distance from the source of the absorber that generates the He$_{\rm \alpha}$ line of Fe XXV. 
The black and red bands are determined by the degree of photoionization (Equation \ref{eq:cond1}) and the line strength (Equation \ref{eq:cond2}).
The orange region is also allowed if the absorption line is saturated. 
(b) schematic drawing of a geometry for the scenario of the accretion disk corona.
\label{fig:scheme}}
\end{figure}

The observational results give constraints on the conditions of the absorber that creates the discovered line.
The presence of $\mathrm{Fe_{XXV}}$ and the absence of the higher ions, $\mathrm{Fe_{XXVI}}$, 
limit the ionization parameter $\xi=L_X/nR^2$ between $\log_{10} \xi = 2.9$--$3.3$ 
(Kallman \& MaCcray \citeyear{Kallman_1982} and Kallman, \citeyear{Kallman_2010}), 
where $L_X$, $n$, and $R$ denote the X-ray luminosity, the number density, and the distance from the ionizing source, respectively.
Assuming $L_X=1\times 10^{38}\;\mathrm{erg\;s^{-1}}$ and $\log_{10}\xi=3.1$, the number density should be 
\begin{equation}\label{eq:cond1}
    n = \frac{L_X}{\xi R^2} = 7.9 \times10^{14} \left(\frac{R}{10^{10}\;\mathrm{cm}}\right)^{-2}\;\mathrm{cm^{-3}}.
    \label{n and R}
\end{equation}
Another constraint is given by the line optical depth:
\begin{equation}\label{eq:tau}
    \tau = N_\mathrm{H} f_\mathrm{Fe} f' \sigma_{\rm lu},
\end{equation}
where $N_\mathrm{H}$ is the hydrogen column density, 
$f_\mathrm{Fe}=2.95 \times 10^{-6}$ is the chemical abundance of iron relative to hydrogen in SMC (Lodders \citeyear{Lodders_2003}, Mucciarelli \citeyear{Mucciarelli_2014}),
$f'=0.5$ is the ion fraction of $\mathrm{Fe_{XXV}}$ among the all charge states of iron (Kallman \& MaCcray \citeyear{Kallman_1982} and Kallman, \citeyear{Kallman_2010}), 
and the cross section at the line center is written by
\begin{equation}
    \sigma_{\rm lu} = 2\pi \sqrt{\pi} r_{\rm e} \hbar c f_{\rm lu} \frac{1}{\Delta E}
\end{equation}
where $r_{\rm e}$, $f_\mathrm{lu}$, and $\Delta E$ are the classical electron radius, 
the oscillator strength of the line transition, and the Doppler width of the line, respectively \citep{Rybicki_Lightman}.
The resonance transition has $f_{\rm lu} = 0.72$ \citep{Kaastra_1996} 
and we determined the line width from fit result, $\Delta E = 6.6\times 10^{-2}$~keV. 
Then, Equation~(\ref{eq:tau}) gives $N_\mathrm{H} = 1.4\times 10^{23}\;\mathrm{cm^{-2}}$.
If we relate a characteristic size $L$ of the absorber to the distance to the NS as $L=aR$, we obtain
\begin{equation}\label{eq:cond2}
n=\frac{N_\mathrm{H}}{L}=2.8 \times 10^{14} \left(\frac{R}{10^{10}\;\mathrm{cm}}\right)^{-1}\;\mathrm{cm^{-3}}.
\end{equation}
The absorber should satisfy those conditions of the density given by Equations (\ref{eq:cond1}) and (\ref{eq:cond2}), both of which are independently derived from different aspects of the observation.
Figure~\ref{fig:scheme}(a) shows these conditions of the absorber in $n$--$R$ space. 
If the absorption line is saturated (see \S \ref{Sec. 3}), regions above the density described by Equation \ref{eq:cond2} is allowed, as shown in Figure \ref{fig:scheme}(a). 

The most probable origin of the absorber is a corona above the accretion disk, or a so-called accretion disk corona (ADC).
Figure \ref{fig:scheme} shows a schematic picture of this scenario. 
This picture of the ADC is supported by the fact that we have observed the absorption line only in a single observation when the NS is at a rising phase of the super-orbital modulation, which can be interpreted as an egress of the disk occultation.
It was at this time that the corona was located in the line of sight, and was able to display the absorption line.
Although the nature of the ADC is poorly understood, it may arise from the disk surface that is illuminated and heated by the strong X-ray radiation from the central accretion flow.
Dense clumps at outer disk, in which the degree of ionization is low, should be responsible for the absorption of the broadband X-rays particularly at low energies, which we observed as short-term dips in the light curves (Fig.\ref{fig:lc}b).
The redshift of $4000_{-1300}^{+1400}\;\mathrm{km\;s^{-1}}$ suggests that the corona moves inward to the NS possibly together with the accretion disk.
Based on the rough estimation of Figure \ref{fig:scheme}(a), 
the distance to the absorber from the NS is smaller than $\sim 10^{11}~{\rm cm}$. 
Since the binary separation is $\sim 2\times 10^{12}~\mathrm{cm}$, 
the line absorption seems to occur at a corona above the inner region of the accretion disk.

We consider other possible scenarios.
Naively, the absorption line would be generated by the stellar wind of the B-type donor star.
However, the absorption line should be blue-shifted since the wind always blows to the observer.
Moreover, the number density of the stellar wind is written as 
\begin{equation}
    n_{\rm wind} = \frac{\dot M_{\rm wind}}{4\pi R^2 v_{\rm wind} \mu m_{\rm p}} \label{wind n}, 
\end{equation}
where $\dot M_{\rm wind} $ is the mass loss rate via the wind, 
$v_{\rm wind}$ is the wind velocity, $\mu$ is the averaged atomic mass number, and $m_{\rm p}$ is the proton mass.
This gives $n_{\rm wind}=4 \times 10^{9}~{\rm cm^{-3}}$ by assuming
$v_{\rm wind} = 1500\;\mathrm{km\;s^{-1}}$, 
$\dot M_{\rm wind} = 1 \times 10^{-6}\;M_\odot\;\mathrm{yr^{-1}}$, and $R = 2\times 10^{12}$~cm.
This number density is too small to meet the conditions of the absorber (see Figure \ref{fig:scheme}a),
and therefore the blowing stellar wind can not explain the observed depth of the absorption line.

%＜第５段落：その他の可能性＞%
%　　・星風捕獲説
%　　　　- 必要な密度とサイズ
%　　　　- ただ、これで説明するためには、星風の密度を**倍くらい大きくする必要がある。
%　　・降着流説
%　　　　- 他の時期では見えない

The wind density can be enhanced by the wind-fed accretion, 
in which the stellar wind is captured by the gravitational potential of the NS.
This is the main accretion mode of the low-luminosity accreting pulsars while SMC X-1, 
which is categorized into high-luminosity systems, accretes matter dominantly via Roche-Lobe overflow.
However, the wind-fed accretion can occur in this system, 
and a red-shifted absorption line can be generated by the captured wind at the orbital phase of 0.5.
A part of the wind within the radius $R_{\rm cap}$ was captured by the potential,
where $R_{\rm cap}$ depends on the NS mass $M_{\rm NS}$ and the wind velocity $v_{\rm wind}$ as
\begin{equation}
    R_{\rm cap} = \frac{2GM_{\rm NS}}{v_{\rm wind}^2}
\end{equation}
where $G$ denotes the gravitational constant. 
Assuming $v_{\rm wind} = 1500~\mathrm{km\;s^{-1}}$, the required number density 
to explain the observed feature 
at the distance of $R_{\rm cap} = 1.3 \times 10^{10}$~cm is estimated to be $1 \times 10^{15}$~cm$^{-3}$ from equation \ref{n and R}. 
To explain the feature with this scenario, the absorber would be required to have a number density that is six orders of magnitude higher than the wind density calculate above.

Another origin of the absorption would be the columnar accretion flow approaching to the NS.
Since the flow has high density, the plasma could still keep helium-like iron in spite of the strong ionizing radiation very close to the NS.
This scenario, however, expects that the absorption feature does not change with both the orbital and super-orbital phases, 
and we conclude that the absorption is took place at more distant part of the accretion flow.

% $$$$$$$$$$$$$$$$$$$ figure 3 $$$$$$$$$$$$$$$$$$$$
% $$$$$$$$$$$$$$$$$$$ figure 3 $$$$$$$$$$$$$$$$$$$$

\section{Conclusions}
We analyzed X-ray data of SMC X-1 obtained with \textit{Suzaku} on May 9, 2012, 
when the NS luminosity was increasing in the super-orbital modulation, 
which can be interpreted as the egress from the occultation by the extend accretion disk around the NS.
The source had an observed luminosity of $1.1 \times 10^{38}\;\mathrm{erg\;s^{-1}}$ in an energy range of 2--10 keV
while the intrinsic luminosity without the extinction 
by the surrounding disk was mildly super-Eddington at $5 \times 10^{38}\;\mathrm{erg\;s^{-1}}$.
We discovered a statistically significant absorption line in the spectrum together 
with an underlying broadened emission line around 6.5 keV.
This absorption line was represented by a narrow absorption line 
centered at $6.61 \pm 0.03$~keV with the line strength of $5.9_{-1.4}^{+1.7}$, 
naturally interpreted as a resonance line of Fe He$\alpha$ with a redshift 
corresponding to a velocity of $4000_{-1300}^{+1400}~\mathrm{km\;s^{-1}}$.

We attribute the absorption line to a corona above the accretion disk, 
which is illuminated and heated by the intense X-ray radiation from the NS.
This interpretation is supported by the fact that the absorption line has been only observed during the egress of the disk occultation.
From the degree of ionization and the line strength, 
we constrained the density and the characteristic length of the absorber, 
and its distance from the NS is smaller than $ \sim 10^{11}\;\mathrm{cm}$.
These well agree with the properties of the corona.
The redshift of the line clearly shows that corona moves inward to the NS, 
implying the corona falls together with the accretion disk.
We have not found a sign of blue-shifted absorption which would be expected if there is an outflow associated with a super-critical accretion.  

\section{ACKNOWLEDGEMENTS}
We would like to thank Prof.\ K.~Osuga, Dr.\ T.~Kawashima, 
Dr.\ H.~Takahashi for useful discussion. 
Also, we would like to thank Dr.\ T. Mihara for critical comments. 
This work was supported by Grant-in-Aid for JSPS Fellows No.\ 16J05852.

%\bibliographystyle{aasjournal}
%\bibliography{APJ_20180623} 

%\begin{thebibliography}{}
%\expandafter\ifx\csname natexlab\endcsname\relax\def\natexlab#1{#1}\fi
%\providecommand{\url}[1]{\href{#1}{#1}}

\end{document}